%% file: decesare-otranto09.tex
\documentclass{PoS}
\usepackage{natbib}
\usepackage{aastex_hack}
\usepackage{deluxetable}

\title{Searching for 511 keV annihilation line emission from galactic compact objects with IBIS}

\ShortTitle{IBIS electron-positron annihilation line}

\author{\speaker{Giovanni De Cesare}\\
INAF-Istituto di Astrofisica Spaziale e Fisica Cosmica di Roma, via Fosso del Cavaliere 100, I-00133 Roma, Italy \\
Dipartimento di Astronomia, Universita' degli Studi di Bologna, Via Ranzani 1, I40127 Bologna, Italy \\
Centre d'Etude Spatiale des Rayonnements, CNRS/UPS, B.P. 4346, 31028 Toulouse Cedex 4, France \\
E-mail: \email{giovanni.decesare@iasf-roma.inaf.it}}
            
\author{P. Ubertini, A. Bazzano \\
            INAF-Istituto di Astrofisica Spaziale e Fisica Cosmica di Roma, via Fosso del Cavaliere 100, I-00133 Roma, Italy \\
           }
            
\abstract{The IBIS imager on board the INTEGRAL satellite, thanks to the large field of view and good sensitivity, gave us a unique opportunity to search for possible 511 keV point sources either previously unknown or associated to known objects such as X-ray binaries or supernovae. The IBIS sensitivity at 511 keV depends on  the gamma ray detector quantum efficiency at this energy and on the background. Both these quantities have been estimated. Reducing all the available IBIS data up to April 2008 we have produced a 5 years full sky 511 keV map with 10 Ms exposure in the Galactic Center. We did not find any significant signal at this energy. The lack of detection of 511 keV galactic point sources is in agreement with the idea that a significant part of the galactic positrons originates in compact objects and then propagates in the interstellar medium.}

\FullConference{The Extreme sky: Sampling the Universe above 10 keV - extremesky2009,\\
		October 13-17, 2009\\
		Otranto (Lecce) Italy}

\begin{document}

\section{Introduction}
The first detection of positrons from the galactic bulge dates from the seventies, when a balloon experiment by \cite{Johnson1972} showed a gamma ray line at the energy of $476\pm 26$ keV from the Galactic Center. However, due to the low energy resolution, the physical origin of this emission was unclear. A few years later, in 1977, thanks to the advent of the high resolution spectrometers, the radiation was identified as the 511~keV line produced by the electron-positron annihilation. Observations of the galactic annihilation line continued progressively until the present days, with the most recent morphological and spectral study by the spectrometer SPI \citep{Vedrenne2003} on board the INTEGRAL gamma ray observatory, as discussed by Jean in this proceeding

Among all possible interpretations on the origin of Galactic positrons \citep[see e.g.][]{Diehl2009}, the LMXBs population could give a relevant contribution; indeed the spatial distribution of these sources within the Galaxy could explain two observed proprieties of the 511~keV diffuse emission. First, the LMXB are old systems concentrated in the galactic bulge, that is where we detect the main flux of annihilation radiation. The second reason is that the recent SPI results show an asymmetric 511~keV emission that can be reproduced by a spatial distribution consistent with the LMXBs population \citep{Weidenspointner2008,Weidenspointner2008a}. However it is not obvious that the source distribution must be correlated to the diffuse proprieties, as the correlation between the source distribution and annihilation diffuse emission depends on how long the positrons propagate in the interstellar medium before annihilation. In this sense, the $\beta^{+}$ decay from supernovae has been proposed as a full solution to explain the SPI observations \citep{Higdon2009}.

Historically transient 511 keV broad lines were observed in two Galactic point sources. One of these sources is the X-ray binary 1E~1740.7-2942, that is well monitored by INTEGRAL  \citep{Delsanto2007}. Discovered by the {\em Einstein}  soft X-ray telescope \citep{Hertz1984}, this source is situated $\approx 48^{\prime}$ away from the radio source Sgr~A*. On October 13 1990 the SIGMA telescope, aboard the GRANAT space observatory, measured a remarkable feature in its emission spectrum, which appears as a bump, reaching a maximum intensity around 500~keV, followed by a $\approx$ 700~keV cutoff. This very broad line contains almost 50~\% of the energy radiated by the source in the 35~keV-1~MeV interval.  The transient feature appeared clearly during a 13 hours observation and then possibly in two further occasions but at a less significant level, leading to controversial conclusions \citep{Jung1995}. Moreover, among the 9 Ms devoted to the Galactic Centre region during the SIGMA mission, these 3 episodes of 511 keV activity represent a very small fraction of time, pointing to a low duty cycle. A 511~keV feature was also reported in Nova Muscae \citep{Gilfanov1991}.

The IBIS imager on the INTEGRAL satellite gave us the opportunity to search for possible 511~keV point sources associated to known objects as X-ray binaries or supernovae, or new ones. Analyzing 5 years of IBIS data, we searched for possible point like 511~keV sources in a time scale of day, month and year. In the next sections we report the data analysis and the results in terms of flux upper limits, and finally a short discussion.

\section{Data analysis}
\input{exp}
We use in our analysis the events detected by ISGRI, that is the lower energy position sensitive detector in the IBIS coded mask telescope \citep{Ubertini2003}. We estimate that the sensitivity that could be obtained using the higher energy layer PICsIT should be about 2 times better. However the PICsIT instrument response is strongly affected by systematic artifacts due to strong detector disuniformities \citep{Lubinski2008}. The correction of  these effects has not yet been fully implemented in the current PICsIT data analysis software release. For this reason we focused on the ISGRI data, while PICsIT will be considered in the future.   

The data reported in this work has been reduced with the OSA 7.0 software release. The data set is made by all the IBIS data available at in April 2008 when we started the analysis, i.e. about 5 year of observations, from October 17 2002, when INTEGRAL was launched, until April 2007 and the Core program data until April 2008. All the selected data correspond to 39413 science windows\footnote{INTEGRAL/IBIS data are organized in short pointings (science windows) of $\sim$ 2000 s} (ScWs). The data sample had been filtered removing the periods  typically occurring during solar flares,  affected by a strong background or a bad detector response. The maximum exposure of about 10 Ms (Fig \ref{fig:exp}) corresponds to the Galactic Center Region, where the bulk of the positron emission is detected by the SPI spectrometer. The data set includes also high latitude observations, since some 511~keV emission might possibly come from Low Mass X-ray binaries in globular clusters. Moreover, if the emission line detected in our galaxy is due to Dark Matter  annihilations, then one should also detect a 511 keV line from nearby dwarf spheroidals \citep{Boehm2004}. 


\input{fwhm}
We made the IBIS sky mosaics in the 431 -- 471~keV, 491 - 531~keV, 551 -- 591~keV energy bands. The width of these bands takes into account the 511~keV line $FWHM$ (measured by fitting the ISGRI background spectra) distribution among the IBIS pointings  (Fig. \ref{fig:fwhm}).

\section{Results}
We do not detect any significant 511~keV signal in day, month time scales as well  as in the 5 year IBIS all sky map. By the estimation of the ISGRI effective area, we are able to put upper limits on the 511 keV flux from point sources. As this limit depends on the square root of the exposure, the best constraint is achieved in the center region of the Galaxy with an exposure of 10~Ms:

\begin{equation}
S_{2\sigma}(Sgr\, A^*) = 1.6 \times 10^{-4}\,ph\,cm^{-2}\,s^{-1}
\label{eq:sgralimit}
\end{equation}

The flux limits for the hard X ($E > 20\, keV$)  microquasars detected by IBIS are shown in table \ref{tab:microquasars}. A better sensitivity for a sample like that can be obtained with some hypothesis staking all the sources signals, but also with this method we do not get any signal at 511~keV.  In table \ref{tab:gcvis} we show the flux limits in shorter (roughly two months) time scales during the IBIS Galactic center visibility periods.


\input{table_microquasars} 
\input{table_sgra}

\section{Discussion}
We have approached our problem in an empirical way: using all IBIS available data to have the best sensitivity and search possible excess at 511~keV on any time scale. The lack of any detection at 511~keV from point sources with IBIS is in agreement with the SPI spectral analysis. Indeed the SPI data suggests that the electron-positron annihilation takes place in a warm interstellar medium: the positrons should travel in the Interstellar medium before interacting with electrons and, as a consequence, the gamma ray emission produced by the annihilation must be diffuse, therefore not-detectable by IBIS with the standard analysis. 

The very broad line features that were detected by SIGMA in 1E~1740.7-2942 and Nova Muscae were explained as being due to the electron-positron annihilation in hot pair plasma in the framework developed by \cite{Ramaty1981}. If this would happen again, we should detect this kind of transient phenomena with IBIS. We can actually confirm with our data that these events have a low duty cycle.

For the future, if the next gamma ray mission EXIST \citep{Grindlay2009}  is definitively approved, the 511~keV line sensitivity for point sources should be improved  by a factor 10 or even more. This progress will be achieved thanks to the larger collecting area and detector thickness, possible with the application of technological advancements in solid state detectors.

\newpage
\bibliography{biblio}{}
\bibliographystyle{aa}

\end{document}

%% file: exp.tex
\begin{figure*}[t]
\centering
\includegraphics[width=14cm,height=6cm]{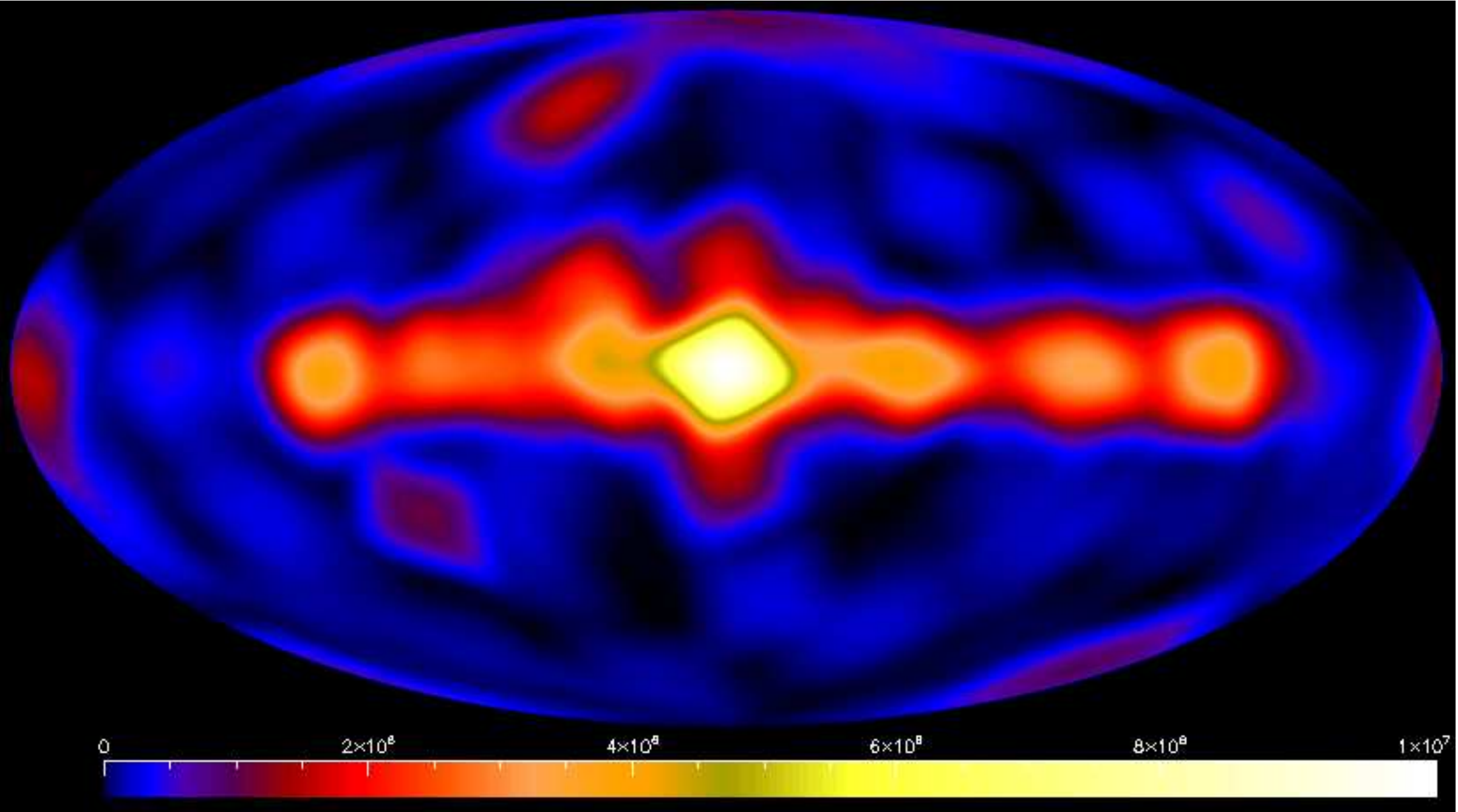}
\caption{The exposure map for our data set. The deeper observation correspond to the galactic plane and the Galactic Center, where we reach with our data a 10 Ms exposure}
\label{fig:exp}
\end{figure*}

%% file: fwhm.tex
\begin{figure*}
\centering
$\begin{array}{cc}
\includegraphics[width=7.5cm,height=5cm]{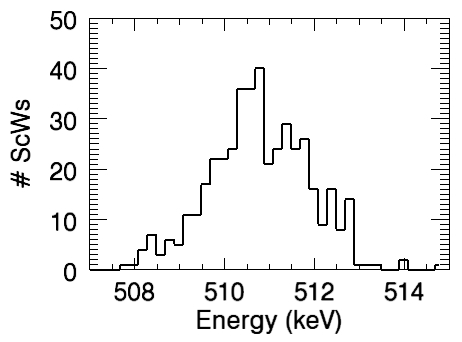} &
\includegraphics[width=7.5cm,height=5cm]{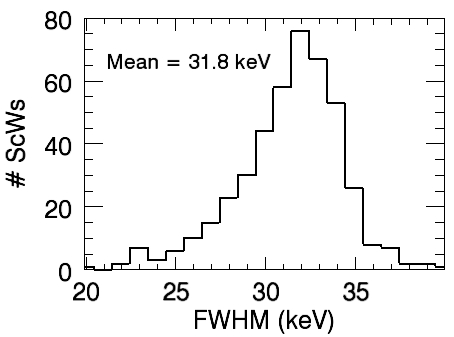}
\end{array}$
\caption{The IBIS/ISGRI response at 511~keV obtained from the analysis of the 511~keV background line: energy (left plot) and FWHM histograms (right plot). The data sample that has been used for these plots  goes from the INTEGRAL satellite revolution 46 to 60}
\label{fig:fwhm}
\end{figure*}

%% file: table_microquasars.tex
\begin{deluxetable}{l r r c r r c c}
\tabletypesize{\footnotesize}
\tablecaption{511 keV $2\,\sigma$ flux upper limits for the hard X (E > 20~keV) microquasars detected by IBIS. These limits have been obtained with the exposure reported in Figure 1.}
\tablewidth{0pt}

\tablecolumns{7}

\tablehead{

\colhead{Source name}  & 
\colhead{R.A.} & 
\colhead{Dec} &
\colhead{Error\tablenotemark{a}} &
\colhead{$F_{20-40}$\tablenotemark{b}} &
\colhead{$F_{40-100}$\tablenotemark{b}} &
\colhead{$2\,\sigma$ flux limit} &
\colhead{Type\tablenotemark{c}} \\

\colhead{} & 
\colhead{(deg)} & 
\colhead{(deg)} &
\colhead{} &
\colhead{(mCrab)} &
\colhead{mCrab} & 
\colhead{($10^{-4}\,ph\,cm^{-2}\,s^{-1}$)} &
\colhead{} 

}\startdata
XTE J1550-564  & 237.745 & -56.479  & 0.2 & $34.0 \pm 0.1$  & $55.3 \pm 0.2$ & 2.7 & LMXB \\       

Sco X-1        & 244.980 & -15.643  & 0.2 & $685.7 \pm 0.3$ & $24.7 \pm 0.3$ & 3.7 & LMXB  \\        

GRO J1655-40   & 253.504 & -39.846  & 0.6 & $2.3 \pm 0.1$   & $2.7 \pm 0.2$  & 2.5 & LMXB  \\        

GX 339-4       & 255.706 & -48.792  & 0.3 & $40.7 \pm 0.1$  & $46.7 \pm 0.2$ & 2.6 & LMXB  \\       

1E 1740.7-2942 & 265.978 & -29.750 & 0.2 & $29.8 \pm 0.1$  & $36.6 \pm 0.1$ & 1.6 & LMXB  \\       

GRS 1758-258   & 270.303 & -25.746  & 0.3 & $58.8 \pm 0.1$  & $75.3 \pm 0.1$ & 1.6 & LMXB  \\         

SS 433         & 287.956 & 4.983    & 0.5 & $10.4 \pm 0.1$  & $5.2 \pm 0.2$  & 2.4 & HMXB  \\      

GRS 1915+105   & 288.799 & 10.944 & 0.2  & $296.8 \pm 0.1$ & $123.4 \pm 0.2$  & 2.5 & LMXB  \\      

Cyg X-1        & 299.590 & 35.199   & 0.2 & $763.7 \pm 0.2$ & $876.7 \pm 0.3$  & 3.0 & HMXB  \\       

Cyg X-3        & 308.108 & 40.956   & 0.2 & $196.5 \pm 0.2$ & $78.3 \pm 0.3$   & 2.6 & HMXB \\     
 \enddata

\tablenotetext{a}{Position error expressed as radius of 90 \% confidence circle in arcminutes}
\tablenotetext{b}{Fluxes averaged on the total exposure \citep{Bird2007}: \\ 
      20-40 keV: $1\,mCrab = 7.57 \times 10^{-12}\,erg\,cm^{-2}\,s^{-1} = 1.71 \times 10^{-4}\,ph\,cm^{-2}\,s^{-1}$ \\
      40-100 keV: $1\,mCrab = 9.42 \times 10^{-12}\,erg\,cm^{-2}\,s^{-1} = 9.67 \times 10^{-5}\,ph\,cm^{-2}\,s^{-1}$}
\tablenotetext{c}{Type identifiers: LMXB=Low Mass X-ray binary, HMXB=High Mass X-ray binary}
\label{tab:microquasars}
\end{deluxetable}

%% file: table_sgra.tex
\begin{deluxetable}{l c c c c c}
\tabletypesize{\footnotesize}
\tablecaption{511 keV $2\,\sigma$ Sgr A* upper flux limit in the Galactic Center visibility periods.}
\tablewidth{0pt}

\tablecolumns{6}

\tablehead{

\colhead{Period}  & 
\colhead{Start Time} & 
\colhead{End Time} &
\colhead{No. ScWs\tablenotemark{a}} &
\colhead{Sgr A* Exposure\tablenotemark{b}} &
\colhead{Flux limit} \\

\colhead{} & 
\colhead{(UT)} & 
\colhead{(UT)} &
\colhead{} &
\colhead{(Ms)} &
\colhead{($10^{-4}\,ph\,cm^{-2}\,s^{-1}$)} 

}\startdata
 2003 Spring & 2003-02-28 & 2003-04-23 & 1731 & 0.650 & 5.7 \\
 2003 Autumn & 2003-08-10 & 2003-10-14 & 1717 & 1.872 & 3.2 \\
 2004 Spring & 2004-02-16 & 2004-04-20 & 1862 & 1.046 & 4.5 \\
 2004 Autumn & 2004-08-17 & 2004-10-27 & 2191 & 1.292 & 4.3 \\
 2005 Spring & 2005-02-16 & 2005-04-28 & 2144 & 1.165 & 4.7 \\
 2005 Autumn & 2005-08-16 & 2005-10-26 & 1667 & 0.790 & 5.9 \\
 2006 Spring & 2006-02-09 & 2006-04-25 & 1869 & 1.501 & 4.1 \\
 2006 Autumn & 2006-08-16 & 2006-11-02 & 1770 & 1.080 & 5.0 \\
 2007 Spring & 2007-02-01 & 2007-04-22 &  985 & 0.347 & 9.2 \\
 \enddata
\tablenotetext{a}{Each ScW lasts about half an hour}
\tablenotetext{b}{The esposure depends both on the number of ScWs and the dithering patterns}
\label{tab:gcvis}
\end{deluxetable}